\documentclass{acm_proc_article-sp}
\usepackage[utf8]{inputenc}
\usepackage{colortbl}
\usepackage{graphicx}
\usepackage{subfigure}
\usepackage{float}
\usepackage{enumitem}

\newcommand{\superscript}[1]{\ensuremath{^{\textrm{#1}}}}
\def\sharedaffiliation{\end{tabular}\newline\begin{tabular}{c}}

\begin{document}
\title{Does query performance optimization lead to energy efficiency?}
\subtitle{A comparative analysis of energy efficiency of database operations
under different workload scenarios}

\numberofauthors{3}
\author{
% First author
\alignauthor
\superscript{1,2}Raik Niemann\\
% Second and third author
\alignauthor
\superscript{2}Nikolaos Korfiatis, \superscript{2}Roberto Zicari\\
% Fourth author
\alignauthor
\superscript{1}Richard Göbel\\
\sharedaffiliation
\begin{tabular}{ccc}
\superscript{1}\affaddr{Institute of Information Systems} & &
\superscript{2}\affaddr{Chair for Database and Information Systems}\\
\affaddr{University of Applied Science Hof} & &
\affaddr{Institute for Informatics and Mathematics}\\
\affaddr{Hof, Germany} & & \affaddr{Goethe University Frankfurt}\\
& & \affaddr{Frankfurt am Main, Germany}\\
\end{tabular}
}

\maketitle

\begin{abstract}
With the continuous increase of online services as well as energy costs, energy
consumption becomes a significant cost factor for the evaluation of data center
operations. A significant contributor to that is the performance of database
servers which are found to constitute the backbone of online services. From a
software approach, while a set of novel data management technologies appear in
the market e.g. key-value based or in-memory databases, classic relational
database management systems (RDBMS) are still widely used. In addition from a
hardware perspective, the majo\-ri\-ty of database servers is still using 
standard magnetic hard drives (HDDs) instead of solid state drives (SSDs) due to 
lower cost of storage per gigabyte, disregarding the performance boost that 
might be given due to high cost.

In this study we focus on a software based assessment of the energy consumption
of a database server by running three different and complete database workloads
namely TCP-H, Star Schema Benchmark -SSB as well a modified benchmark we have
derived for this study called W22. We profile the energy distribution among the
most important server components and by using different resource allocation we
assess the energy consumption of a typical open source RDBMS
(\textit{PostgreSQL}) on a standard server in relation with its performance
(measured by query time).

Results confirm the well-known fact that even for complete workloads,
optimization of the RDBMS results to lower ener\-gy consumption.
\end{abstract}

\newpage


\section{Introduction}
A general assumption is that efficient programs are also ener\-gy
efficient\cite{/23/}. This seems to be reasonable since a program requiring
less computation time will probably also require less energy for its execution.
This can be true if we assume a constant energy use of a computer, and the
computer performs more tasks in the same amount of time because of time
efficient programs.

However, a more detailed analysis may show a different picture. For example it
is a well-known fact that one complexity measure like time can be optimized at
the costs of another complexity measure of space. An optimization following this
path may require that the increased amount of data for a program may be
transferred from one type of memory to a different type of memory. In a modern
computer this might result that the data may be moved from the CPU cache to the
main memory or from main memory to a drive. An obvious consequence of this
transfer are longer access times which may already reduce the benefit of these
optimization strategies. In addition the usage of a different memory type may
result in higher energy costs\cite{/7/}. For example, this can be the case if
a large data structure needs to be generated on an external drive and the access
to this drive would need to be frequently used in comparison to a solution where
the drive may be even stopped to save energy.

Since the amount of information to be processed has drastically increased over
the last couple of years, large and very large data centers which are expected
to support internet operation have been built by major players (e.g.
\textit{Google} and \textit{Amazon}). However, increased costs of operation
(electricity, cooling costs) have made this vendors to offer the rental of
unneeded data center capacity (disk storage space, computation time) to nearly
everyone who is willing to pay. This gives rise to the concept of
``\textit{cloud computing}'' where, for example, server capacities are only
needed for a limited time or where the acquisition of an own data center is too
expensive\cite{/18/}.

Furthermore, in the context of database applications another issue might arise:
the distribution of data across a database cluster as a basis for distributed
computing, as it happens in the case of Big Data/\textit{Hadoop}
clusters\cite{/17/}. Industry trends\footnote{\textit{State of the data center
2011} by \textit{Emerson}} advocate that increased demand for server performance
will be fulfilled by deploying more (database) servers and data centers. But
with increased energy costs, data centers ope\-ra\-tors are looking for 
mechanisms to avoid or reduce those costs (scale-up strategies). This way the 
energy efficiency of a single database server becomes important as it can be a 
crucial component for the overall cost assessment of a data center operation 
where energy costs are the most significant factor on their operation and
scalability. 

\subsection{Background and motivation}
Several approaches which can be found in the literature have addressed the issue
of energy efficiency of database servers from different aspects.
\textit{Harizopoulos et. al}\cite{/3/} as well as \textit{Graefe}\cite{/26/}
divided energy efficiency improvements into a hardware and a software part.
Recent work such as the one by \textit{Baroso} and \textit{Hölzle}\cite{/19/}
has revealed the fact that the impact of hardware improvements on the energy
efficiency is quite low. A main approach in that direction is the use of Solid
State Drives (SSD's) which have been shown to improve performance on typical
database operations such as sorting and therefore energy
efficiency\cite{/14/}. Obviously SSDs are faster in such kind of
scenarios\cite{/5/}, but there are other tradeoffs as a replacement for HDDs
with most obvious the cost per gigabyte (\textit{Schall et. all}\cite{/1/},
\textit{Schröder-Preikschat et. all}\cite{/2/}, \textit{Härder et.
all}\cite{/27/}).

On the other side, software improvements seem to be more attractive when it
comes to increasing the energy efficiency of a database server, especially for
relational databases. \textit{Lang} and \textit{Patel}\cite{/13/}, for example,
proposed a database query reordering technique to influence the energy
consumption of the database server. \textit{Xu et. all}\cite{/10/} proposed a
modification of the query planner in order to take the estimated energy
consumption into consideration.

\textit{Tsirogiannis et. all}\cite{/7/} had a very detailed investigation
regarding the relation between performance and energy consumption in a RDBMS.
For that they did extensive measurements with various hardware configurations
that are typically found in a database server for a scale-out scenario. In
addition to this, they identified performance tradeoffs for different SQL query
operators.

\subsection{Objective of this study}
This study provides the basis for a more detailed analysis of the energy
consumption in the context of database applications that are common in daily
workloads of enterprise users. For this purpose the paper does not only consider
the total power consumption of the full system but also the usage of energy by
different components. In addition the paper analyses standard energy saving
strategies coming with a modern computer system and their impact on the power
consumption.

In the context of database applications, it is important to know which database
operation or which combination of SQL operators affects which of the measured
components. This study analyzes these combinations and their fraction of the
overall power consumption.

All the mentioned aspects lead to the final view on the energy effiency of a
single database server. This is important when it comes to a scale-out scenario
that is typically found in a database cluster. There are two choices: either one
decreases the energy consumption of the used hardware equipment with a small
decrease of the database response time, or increases the performance while
accepting a slight increase of the energy consumption. Both ways improve the
energy effiency.

As suggested by \textit{Xu}\cite{/28/}, more experimental research has to be
done concerning the energy consumption of database servers. Taking this into
account, this study tries to have a closer look on both choices mentioned
above and to give recommendations how to improve the energy effiency of a single
database server in a general way (for example the usage of buffers and indices
combined with traditional HDDs as the primary storage for the database files).

\section{Measurement methodology}

\subsection{Test server preparation}
For the purpose of this study, we constructed a database server making use of
recent technical core components. The operating system selected for testing was
a typical \textit{Linux} distribution (\textit{Ubuntu Server} version 11.10) 
using a stable kernel (kernel version: 3.0.0.17). In order to eliminate biases 
from the operating system in our measurements, all unnecessary operating system 
services were turned off. The database management system (DBMS) that was used 
was \textit{PostgreSQL} version 9.1. The hardware characteristics of the test 
server are provided in table \ref{tab::configuration}.

\begin{table}[H]
\small\centering
\begin{tabular}{|l|p{.6\columnwidth}|}
\hline CPU & \emph{Intel Core i7-860 @2.8 GHz} \\
\hline Main memory & 3x \emph{Samsung} DDR-2 2 GByte 800 MHz
(m378b5673fh0-ch9)\\
\hline Hard drives & 3x \emph{Hitachi} 1 TByte, 16 MByte cache,
(HDT721010SLA360) \\
\hline
\end{tabular}
\caption{Hardware characteristics of the test server used in our
study}\label{tab::configuration}
\end{table}

Two of the three hard drives were combined as a striping RAID array in order to
boost performance and to separate the filesystem calls of the DBMS from the
operating system. The operating system itself was installed on the remaining
third hard drive.

We identified the core components we were interested in their power consumption
throughout the various tests as follows: (a) CPU, (b) main memory, (c)
motherboard as a whole and (d) the hard drives of the RAID array. These core
components were used as a unity of analysis in order to assess the optimization
options. Additionally we were interested in the impact of the operating system
settings as well as the test server capabilities on the energy consumption of
the core components. Appendix \ref{sec::testarrangement} provides the complete
test arrangement.

Besides, the test server's motherboard used \textit{Intel}'s 
EIST\footnote{Acrynym for \textit{Intel Enhanced Speedstep}. It allows the 
system to dynamically adjust the CPU frequency and voltage to decrease energy 
consumption and heat production.}. In general \textit{EIST} enables and disables 
CPU cores or reduces and raises the overall CPU clock frequency as a function of 
the CPU usage. This also affects other technical aspects, e.g. the heat 
dissipation from the CPU.

Recent Linux kernels offer several modules providing more data to the frequency
scaling heuristics. The most handful modules for our test server configuration
are the \textit{on-demand} and the \textit{power-saving} modules. Taking this
into account, the command line tool \textit{powertop}\footnote{Refer to
\texttt{http://www.lesswatts.org/projects/powertop/}} was used to suggest
configurations on the disablement of operating system services that might
influence energy consumption.

\subsection{Stress tests on energy consumption}
\begin{figure}[htb]
\includegraphics[width=\columnwidth]{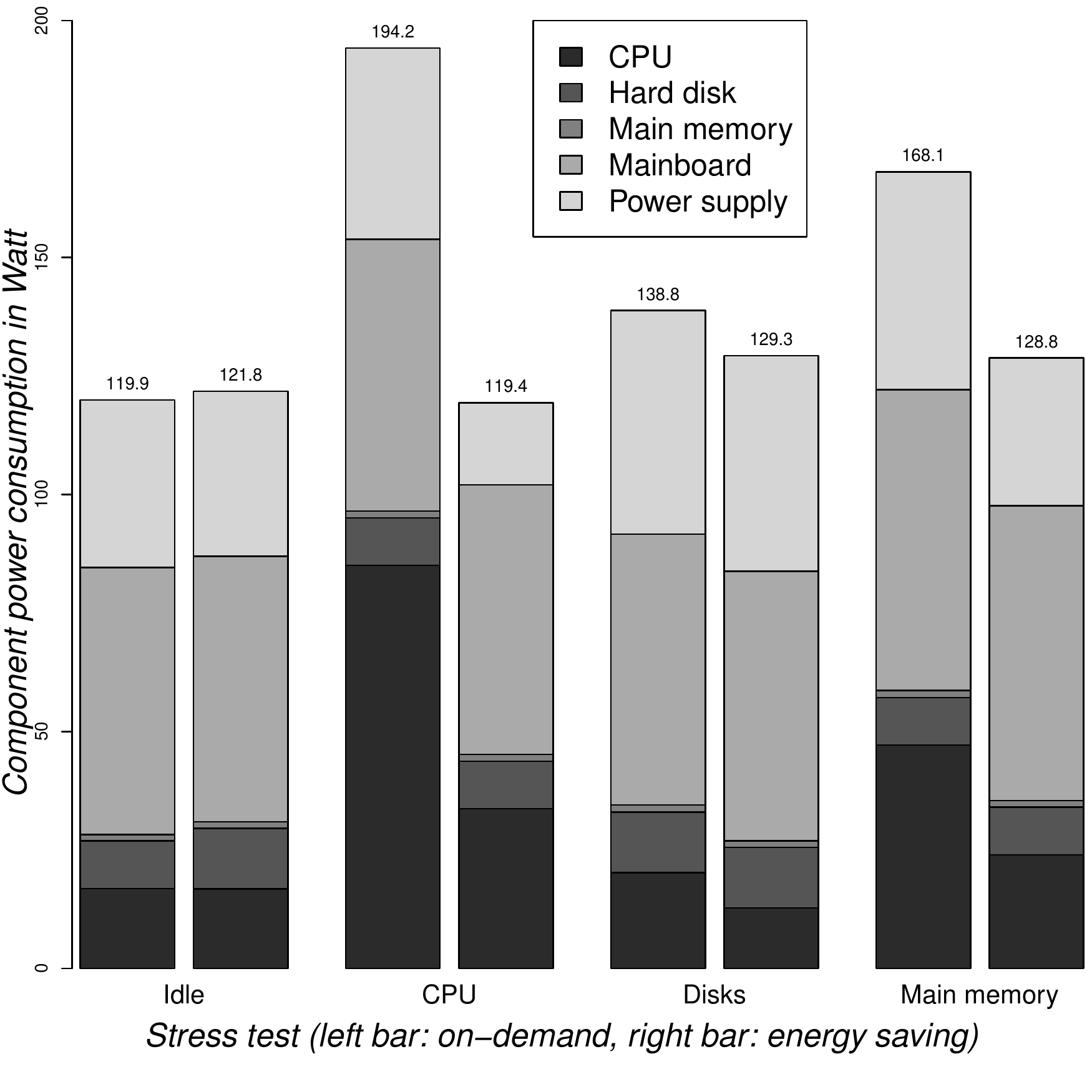}
\caption{Energy consumption per component during our calibration stress
test}\label{fig::tests}
\end{figure}

Before assessing the energy efficiency as a whole as well as the components
we were interested in profiling their energy consumption, we performed stress
tests to get a detailed overview of the energy distribution among the single
components. Figure \ref{fig::tests} summarizes the stress tests. Each tuple of
the shown bars represents a stress test for a core component. The left bar of a
specific tuple illustrates the energy consumption of each component in
conjunction with the \textit{on-demand} frequency scaling module. The right bar
of the tuple displays the energy consumption of all energy saving settings
enabled, respectively. Figure \ref{fig::tests} also shows the energy consumption
of the power supply unit (PSU) which is the difference between the overall power
consumption of the test server and the one of the measured components.

Our first action was to calibrate the test procedure by analyzing  the energy
consumption of the core hardware components as a basis for the other tests. All
services and applications not required for the operating system as well as the
DBMS service were turned off. The operating system uses default settings, e.g.
the usage of the \textit{on-demand} CPU frequency scaling module. The average
consumption per core component is shown in figure \ref{fig::tests} in the left
bar of the tuple named \textit{Idle}. The right bar of this tuple shows the
energy consumption with all available energy saving settings turned on. There
is a slight increase of the power consumption: although the CPU is forced to
reach the energy saving idle states, it is constantly interrupted by doing so,
for example by administrative background processes of the operating system.
Changing the CPU state costs some effort but results in higher energy
consumption.

To measure the power consumption stressing the CPU of our test server, we
decided to use the \textit{Linux} command line tool \textit{burnMMX} because it
utilizes all parts of the CPU including extension like \textit{MMX} or
\textit{AVX}. Based on the eight CPU cores reported to the operating system, we
ran our CPU stress test eight times and increased the number of running
\textit{burnMMX} instances accordingly. As assumed the power consumption
increases up to the four real cores and remains relatively constant if more
cores were used due to HT\footnote{\textit{HT} is an acronym for
\textit{Hyperthreading} by \textit{Intel}. It is used to improve
parallelization of computation.}. This behavior corresponds to the CPU tests
reported in \cite{/7/}. The CPU stress test with the highest energy consumption
for both scenarios (energy saving settings turned off and on) is depicted in
figure \ref{fig::tests} as the bar tuple labeled \textit{CPU} for comparison
with the other stress tests.

To stress test the hard drives in the RAID array we executed the I/O benchmark
suite \textit{iozone} while observing the power consumption throughout the
different benchmark tests (different access strategies, buffered and un-buffered
data access and so on). Our test results are displayed in figure
\ref{fig::tests} in the in the second bar tupel labeled \textit{Disks}. Our test
shows that the impact on the energy consumption is low when all energy savings
settings are turned on. However, in contrast to the power consumption of current
SSDs\cite{/2/} the power consumption of the hard drives is two to three times
higher.

Finally we stressed the main memory with the command line tool
\textit{memtester} that uses different patterns to access the main memory. It
also tests for the correctness of the contents by writing, reading and comparing
the main memory areas. The effect on the power lane supplying the main memory
was not measurable. In contrast to this, the left bar of the tuple named
\textit{Main memory} in figure \ref{fig::tests} indicates an increase of the
overall power consumption of 30 Watts in which the CPU is responsible for. Even
with all energy saving settings turned on\cite{/9/}, the overall energy
consumption was higher than in idle state.

\subsection{Measuring energy efficiency}\label{sec::formula}
We define the performance ratio ($P$) of a single database query as the unit of
time to execute the query in time ($t$) to obtain the results\cite{/12/}:
\begin{equation}
P = \frac{1}{t}
\end{equation}
We then normalize a set of performance values to values between zero and one
as follows:
\[
P_{normalized,i} = P_i \cdot \frac{1}{max(P)}
\]
We then define the \textit{energy efficiency} $EE$ as the ratio between the
performance $P$ of the database query and the electrical work $W$
executing the query:
\begin{equation}
EE = \frac{P}{W}
\end{equation}
We then normalize the set of efficiency values to values between zero and
one as follows:
\[
EE_{normalized,i} = EE_i \cdot \frac{1}{max(EE)}
\]

\subsection{Selection of workloads and DBMS parameters}
As aforementioned our intuition here is to examine the effects of executing a
database query in relation to the overall power consumption. For example a
combination of a not well formed SQL query and an optimized query planner can
cause a cascade of operations that consumes a lot of unnecessary power, e.g.
sequential scans cause unnecessary hard drive accesses with all its overhead in
the operating system (access control, swapping and so on).

According to related work, for example \cite{/7/} or \cite{/19/}, we identified
several settings for our \textit{PostgreSQL} database server we hypothesized
they had an important impact on the power consumption. Those settings can be
divided into two groups: the first one are the settings for the underlying
operating system and for \textit{PostgreSQL} and the second one are settings for
the database itself. In detail those settings are: 
\begin{itemize}[topsep=0pt,itemsep=0pt]
\item The size of the main memory assigned to \textit{PostgreSQL} to operate
\item The size of the miscellaneous buffers, e.g. for sorting resulting rows
or caching
\item The settings for the query planner
\item The size of the data in the database
\item The session type of executing subsequent queries (single session vs.
multi session)
\item Combinations of the different SQL operators and functions, for example
increasing number of joined tables or the number of result set dimensions to be
restricted
\end{itemize}
To get comparable results we decided to run three complete database benchmark
workloads: \textit{TPC-H}\cite{/4/}, the \textit{Star schema benchmark} 
(SSB\cite{/22/}) and a third benchmark constructed specifically for this study,
we call it the \textit{W22} benchmark. The first two offer a standardized way
for comparison whereas the last one is a workload we composed to analyze the
behavior of \textit{PostgreSQL} not covered by the previously mentioned
workloads.

\section{Workload results}

\subsection{TPC-H workload}\label{sec::TPC-H}
For the TPC-H workload we used their data generator to generate three databases
with a size of 1, 5 and 10 GByte of data as well as all suggested indices. We
chose this database sizes because we wanted the databases to fit completely,
nearly and under no circumstances in the working main memory for
\textit{PostgreSQL} by assigning 1, 2.5 and 5 GByte. This selection was made due
to the limitation of 6 GByte overall main memory present in the database server.
In a final setup step we optimized the internal data structures and the query
planner statistics of \textit{PostgreSQL} for the created databases.

\begin{figure}[H]
\centering
\includegraphics[width=\columnwidth]{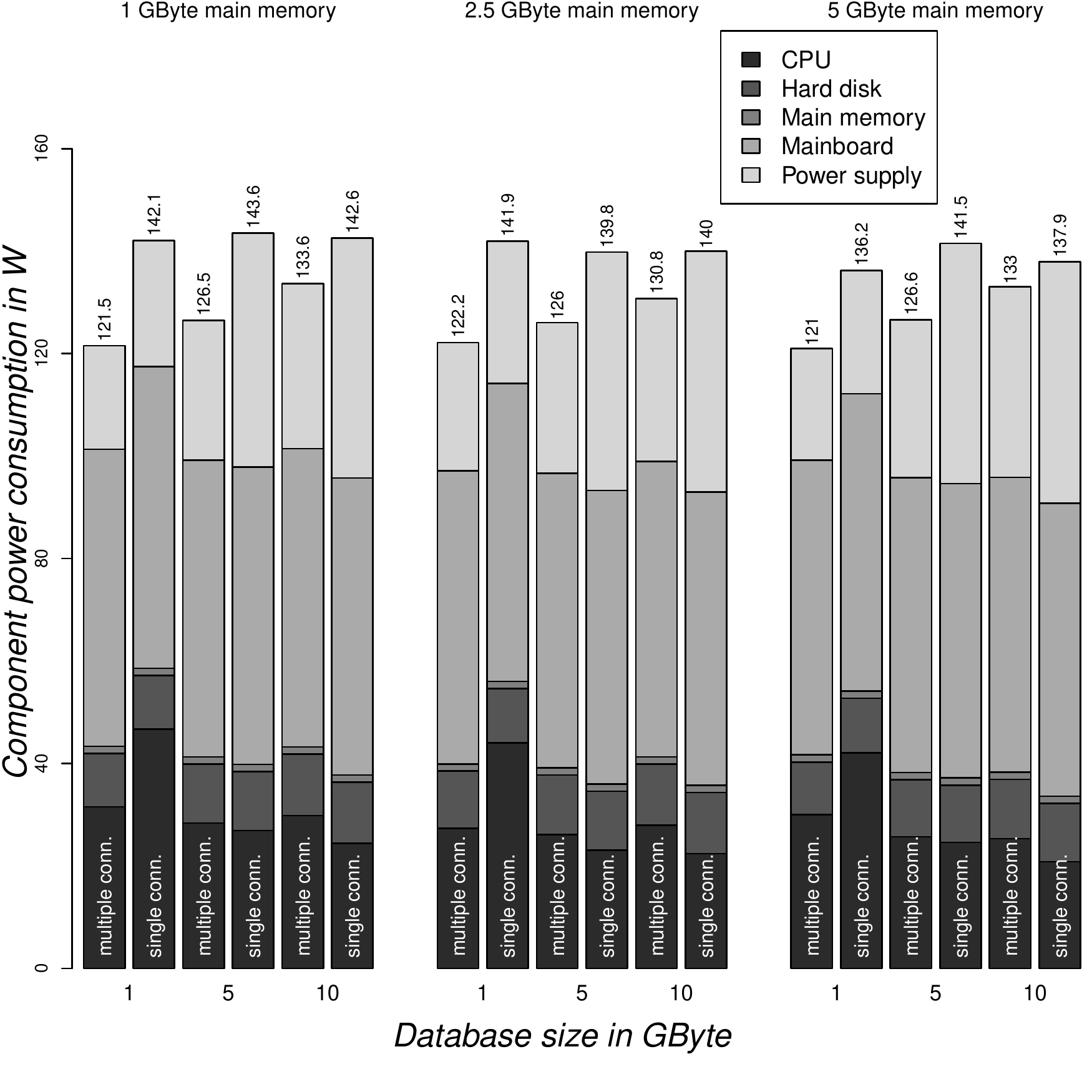}
\caption{Energy consumption per component for TPC-H workload}
\label{fig::tpch-consumption}
\end{figure}

At first we ran the benchmark queries subsequently using a new database
connection to avoid \textit{PostgreSQL}'s internal cache\footnote{Notice that
\textit{PostgreSQL}'s implementation isolates client connections completely by 
running the client sessions in shared-nothing processes. The client processes 
only share some IPC memory for synchronization purposes and the ACID 
functionality. This means that every client has its own cache.}. After this we 
retried the TPC-H benchmark using a single database connection for the SQL 
queries. The average power consumption per component for both test series is 
shown in figure \ref{fig::tpch-consumption}.

\newpage
This figure clearly indicates the higher energy consumption when a single 
session is used. We recognized that the average overall power consumption for 
all of our TPC-H tests does not vary a lot. Please compare only the bars which 
are identically labeled with each other. We assumed that the CPU and hard 
drives 
are the main energy consumer but in fact it turned out that the mainboard is 
the 
biggest one.

\begin{figure}[H]
\centering
\includegraphics[width=\columnwidth]{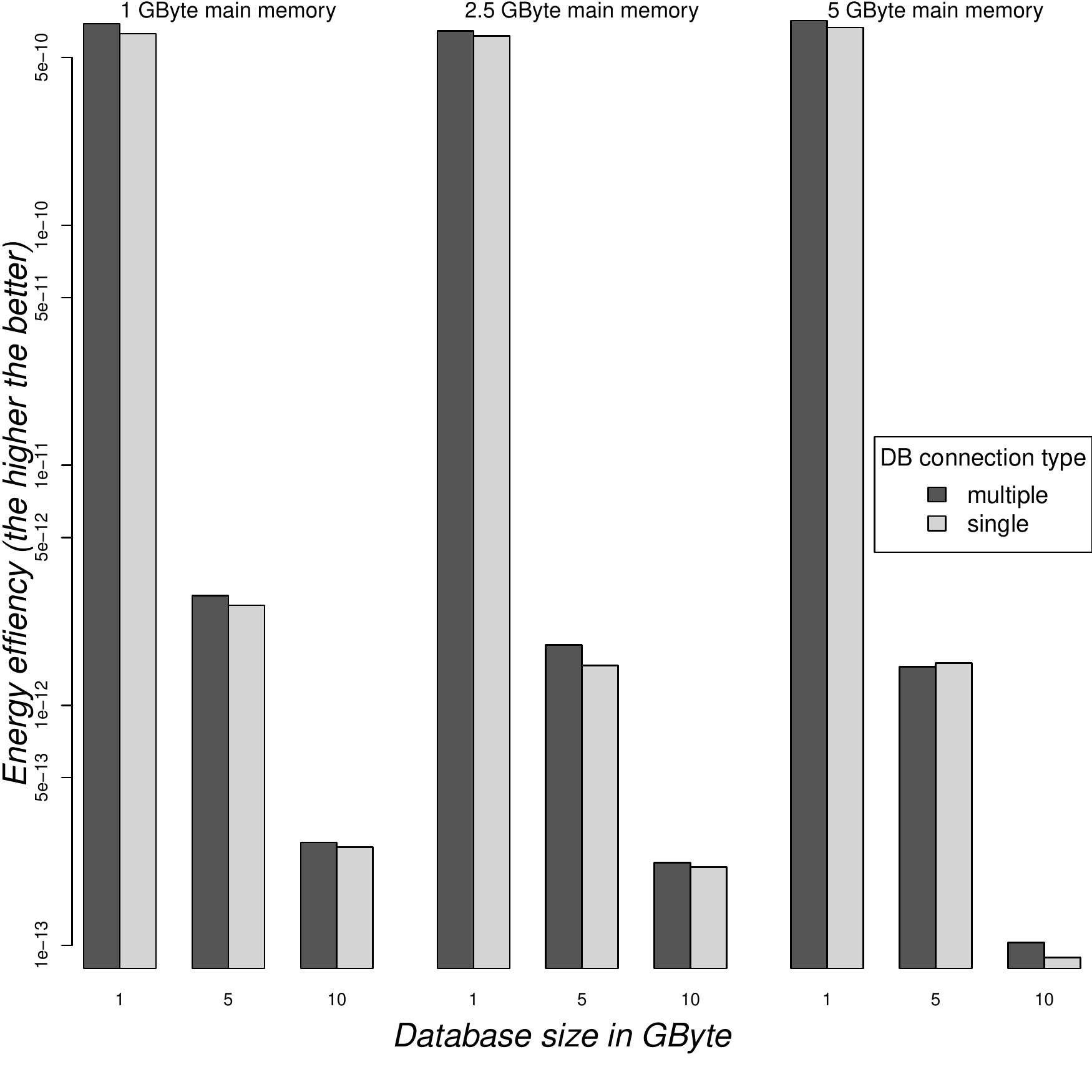}
\caption{Energy effiency for TPC-H workload}\label{fig::tpch}
\end{figure}

We also assumed that the usage of single database connection would improve the
performance because of the better usage of \textit{PostgreSQL}'s internal
cache but the opposite occured: the performance was nearly the same with an
increased power consumption of 7 percent on average. Using the equations
outlined in section \ref{sec::formula}, this leads to a lower energy effiency of
nearly 8 percent on average as depicted in figure \ref{fig::tpch}.

A detailed study of the query plans reveals a broad usage of sequential scans on
the TPC-H database tables as well as a low usage of the internal caches.
Therefore we modified \textit{PostgreSQL}'s settings regarding the query planner
and caches to favor the provided indices and repeated the tests. This has an
inverted effect: the performance decreases by about 4 percent on average. The
reason is the overhead to process the indices which are also disk bounded.

In general we observed a big effect on the energy effiency by assigning a
higher portion of the main me\-mo\-ry to \textit{PostgreSQL}. This results in
massive swapping for some TPC-H queries, e.g. 1, 9 and 21, and the suspension of
the \textit{PostgreSQL} process. In fact, \textit{PostgreSQL} is very disk bound
and by accessing the database files the operating system swaps heavily because
of the reduced main memory portion. Besides, swapping does effect only the hard
drive and not the CPU. This affects the overall power consumption for our TPC-H
tests and explains the small variation of the values.

\begin{figure}[H]
\centering
\includegraphics[width=\columnwidth]{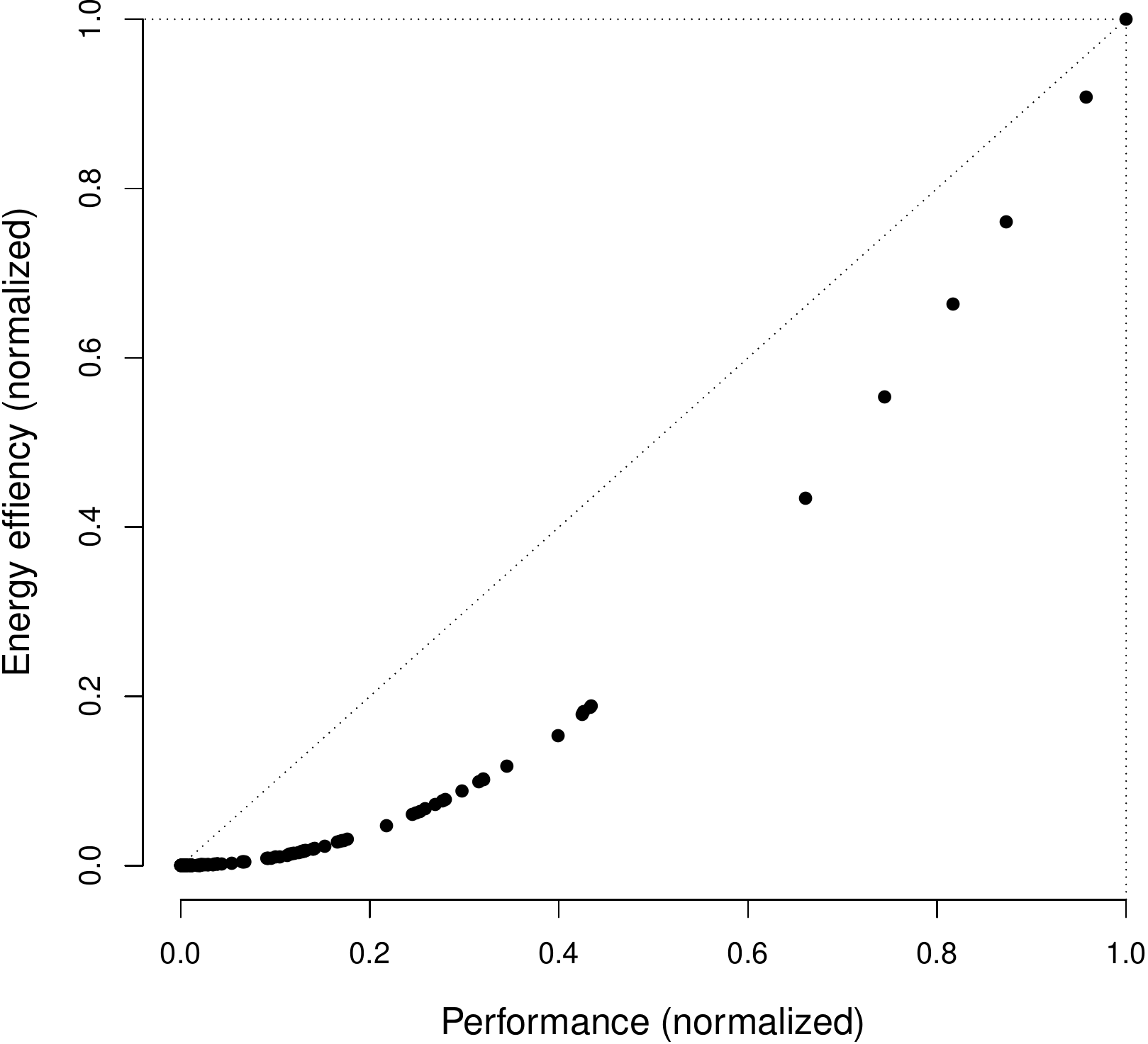}
\caption{Performance vs. energy effiency for TPC-H workload}
\label{fig::normalizedEffiency-TPCH}
\end{figure}

Finally we were  interested in performance versus ener\-gy effiency ratio as
stated in the conclusion of \cite{/7/}. This ratio for our  entire TPC-H tests
with all its different configurations is shown in figure
\ref{fig::normalizedEffiency-TPCH} and confirms the strong relation between
performance and energy effiency (\emph{the most energy-effiencent configuration
is typically the highest performing one}).

This figure also shows that the majority of the configurations are clustered in
the lower left area. This means that most of the queries of the TPC-H benchmark
show a poor energy effiency.

\subsection{Star schema workload}\label{sec::SSB}
The \emph{Star schema benchmark} (SSB) was initially composed to get a database
layout closer to reality compared to TPC-H which it is based on. According to
\cite{/22/} several original tables were decoupled to make many join
ope\-ra\-tions unnecessary and the set of SQL queries of SSB were created to be
more realistic and to test the database capabilities regarding range coverage
and indice usage.

For executing the SSB SQL queries we used the same parameters as described in
section \ref{sec::TPC-H}: we generated three different databases with 1, 5 and
10 GByte of data and created test configurations for \textit{PostgreSQL} with 1,
2.5 and 5 GByte main memory to work on.

\begin{figure}[H]
\centering
\includegraphics[width=\columnwidth]{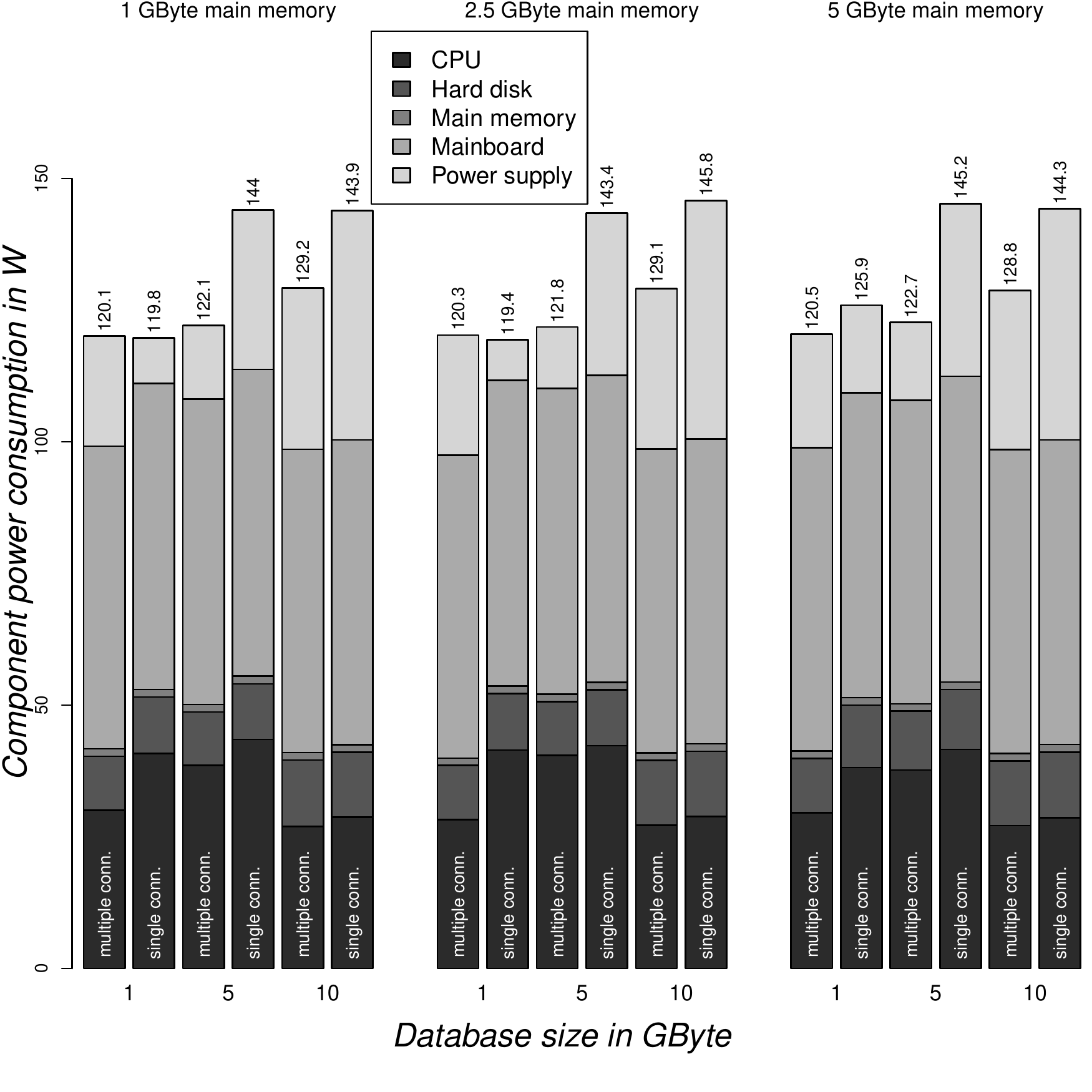}
\caption{Energy consumption per component for the SSB workload}
\label{fig::SSB-consumption}
\end{figure}

Just as for the TPC-H workload, we ran the SSB database queries subsequently by
using a single and multiple database connections. The ave\-ra\-ge energy
consumption per component is depicted in figure \ref{fig::SSB-consumption}. The
type of accessing the database matters: although the ave\-ra\-ge performance is
near\-ly the same, the average power consumption for multiple connections is
lower than the one for using a single connection. The latter consumed nearly 10
percent more energy on average. Except for the database size of 1 GByte, this
results in a lower energy effiency.

\begin{figure}[H]
\centering
\includegraphics[width=\columnwidth]{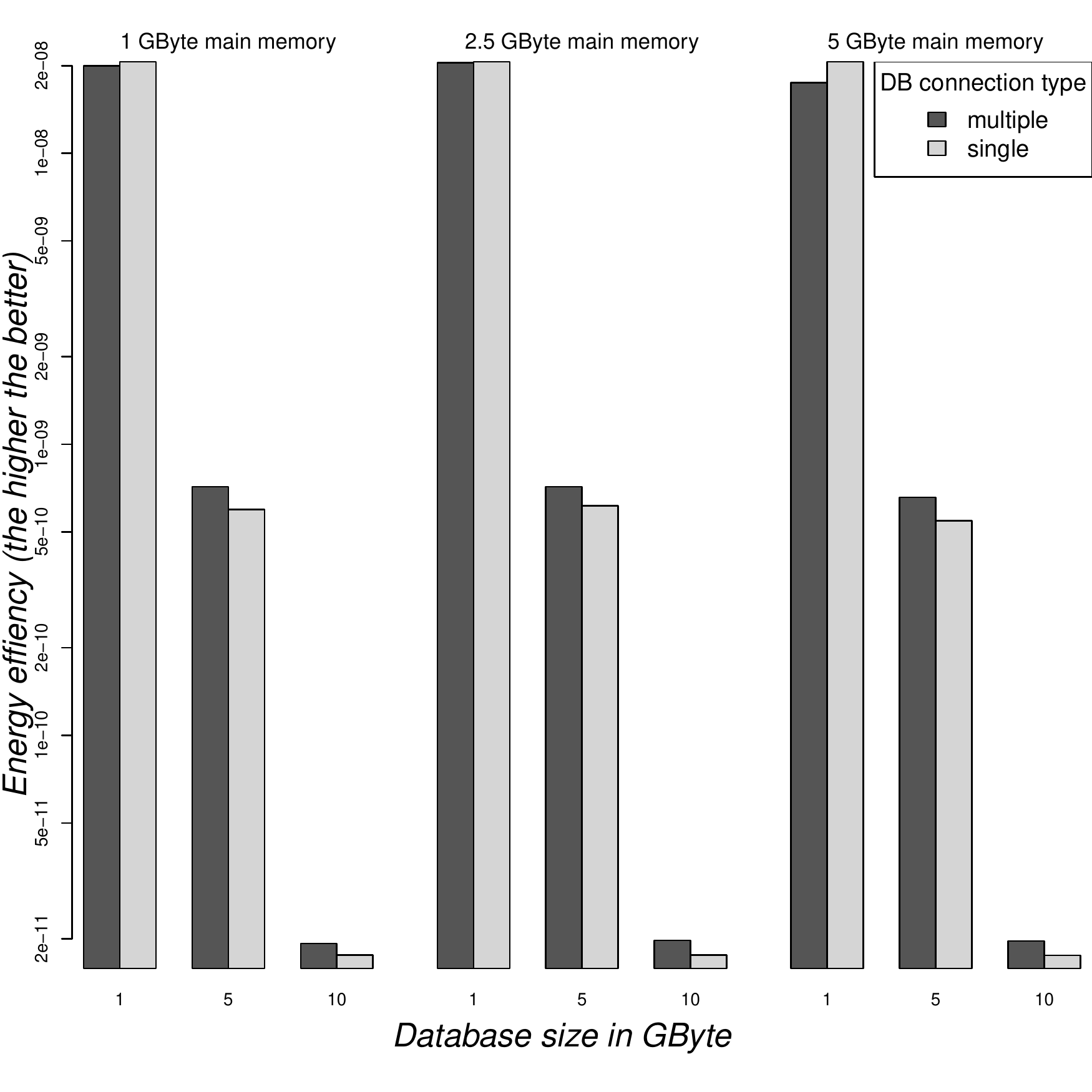}
\caption{Energy effiency for SSB workload}
\label{fig::ssb}
\end{figure}

In general and in terms of energy effiency, the SSB benchmark performs better
than TPC-H. In addition to this, SSB organizes its SQL queries into groups in
which the requested rows do not overlap and the group number indicates on how
many dimensions the result set has to be restricted. This allows a better
analysis of the results and avoids side effects interfering the results. Figure
\ref{fig::ssb} shows the energy effiency of our SSB tests.

A deeper investigation of the query plans showed us that the reduced set of
tables and the modified queries lead to a more stable and predictable behaviour
con\-cer\-ning the power consumption and energy effiency. In particular this
can be seen in figure \ref{fig::ssb}: the energy effiency for a given
constant SSB database size does not vary a lot and is relativly independent from
the amount of main memory spent for \textit{PostgreSQL}.

Besides, the energy effiency of the SSB database with 10 GByte is remarkably
lower than the ones with 1 and 5 GByte of data. An investigation of the logged
system activities during the tests revealed heavy swapping actions which caused
the suspension of the \textit{PostgreSQL} process handling the que\-ries. As
illustrated in figure \ref{fig::SSB-consumption}, the low CPU energy consumption
indicates the heavy swapping activity.

\begin{figure}[H]
\centering
\includegraphics[width=\columnwidth]{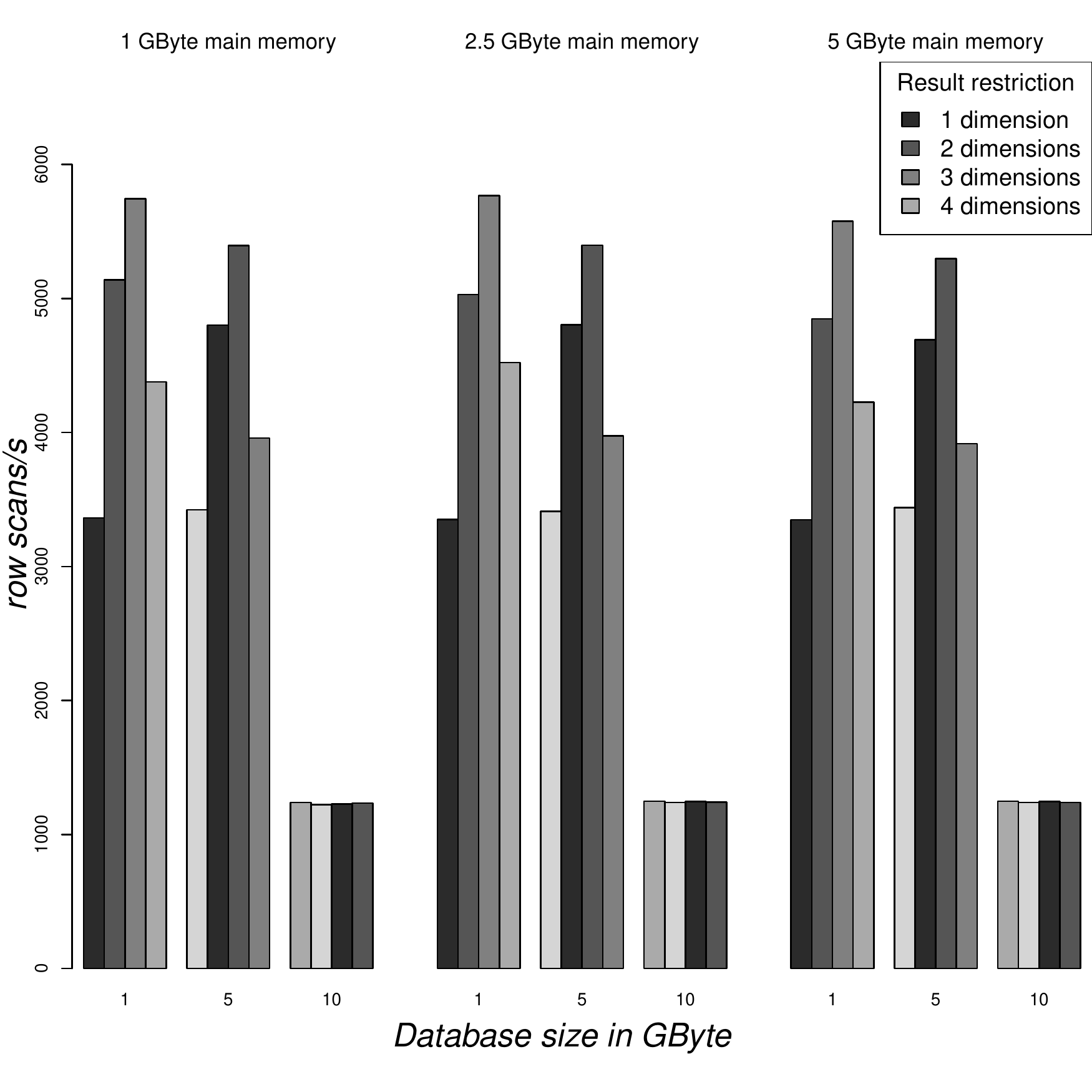}
\caption{SSB row scans per second}
\label{fig::SSB-Rowscans}
\end{figure}

The query plans also show the general use of sequential scans on the SSB
database tables. In fact, no indices were involved to execute the queries in
all configurations. The execution of the queries purely relies on the read
performance of the database files. As figure \ref{fig::SSB-Rowscans} indicates,
it does not matter if the result set of the joint tables are further
restricted. In addition to this, this figure shows that main memory spent for
buffers is relativly unimportant: if one consider a specific database size, the
row scan per second rate does not vary a lot. This means that the mentioned rate
is independent of the main memory fraction assigned to \textit{PostgreSQL} and
also independent of the number of dimensions the result set is restricted.

\begin{figure}[H]
\centering
\includegraphics[width=\columnwidth]{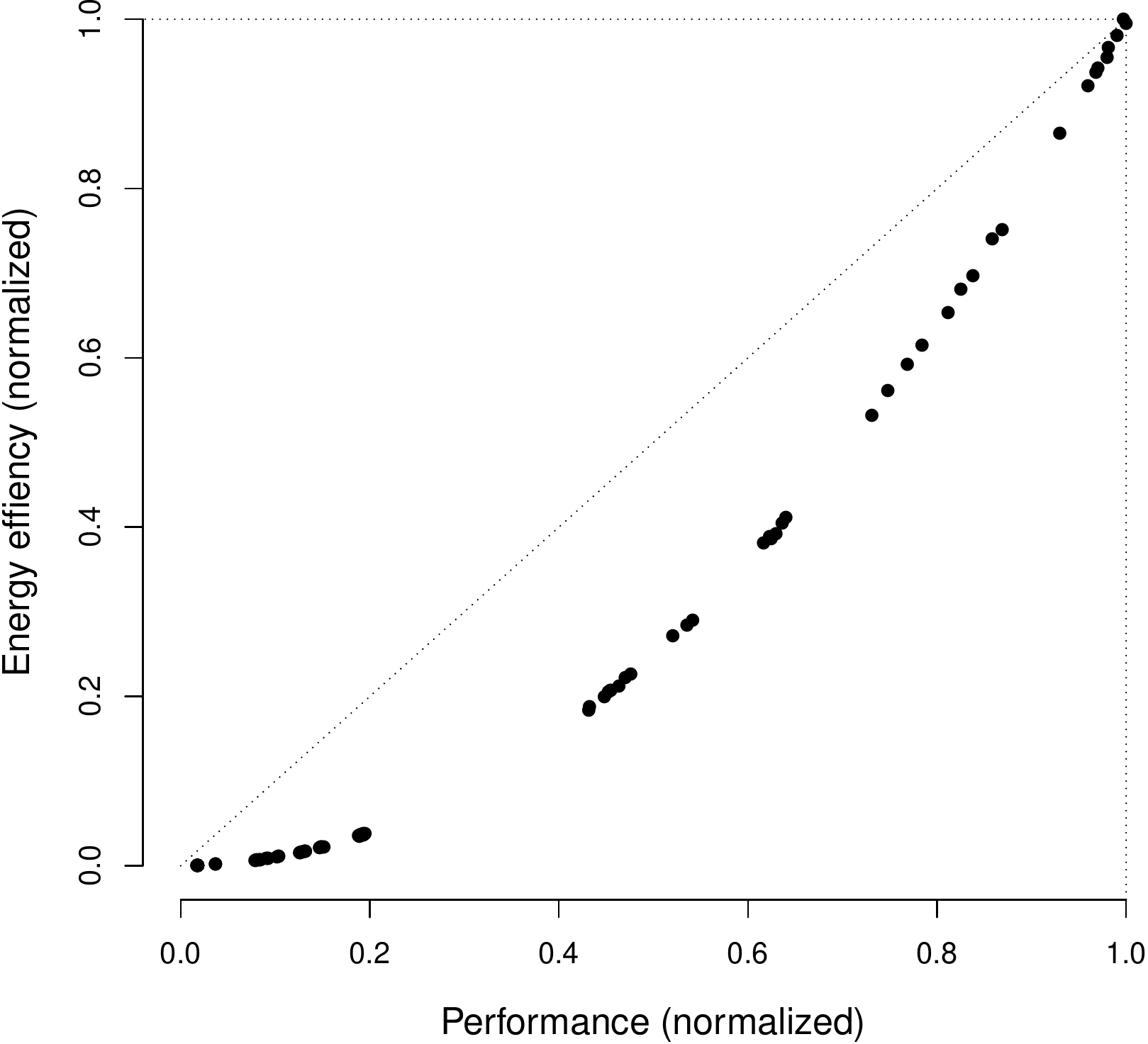}
\caption{Performance vs. energy effiency for SSB}
\label{fig::normalizedEffiency-SSB}
\end{figure}

\newpage
The performance vs. energy effiency ratio for all SSB test configurations is
illustrated in figure \ref{fig::normalizedEffiency-SSB}. The same strong
relationship can be seen as in figure \ref{fig::normalizedEffiency-TPCH} for the
TPC-H benchmark tests. In contrast to TPC-H, the vast majority of the SSB test
configurations is clustered in the upper right corner of figure
\ref{fig::normalizedEffiency-SSB}. This means that the SSB configurations
perform much better. This leads to a better energy effiency.

\subsection{W22 workload}
To get a more in-detailed look at the impact of the different SQL operations
and their combinations we composed a database workload called W22. The workload
consists of six groups of database queries:
\begin{enumerate}[topsep=0pt,itemsep=0pt]
\item aggregate functions (\texttt{count}, \texttt{avg} and \texttt{sum})
\item grouping (\texttt{GROUP BY})
\item sorting (\texttt{ORDERED BY})
\item selecting different data types, e.g. \texttt{int}, \texttt{var\-char},
\texttt{text} and \texttt{date}
\item removing duplicates (\texttt{DISTINCT})
\item joins (cross joins, conditional joins)
\end{enumerate}
In contrast to this, the TPC-H and the SSB workload are designed for benchmark
an OLAP scenario. Their database queries use a mix of different SQL operations
coming from the groups above. Our W22 workload instead tries to analyze the
impact of SQL queries in each group. Besides, this workload allows to analyze
the behaviour of \textit{PostgreSQL}'s internal query optimizer and query
planner as well as the interaction with the underlaying operating system.
The W22 queries operate on the TPC-H databases described in section
\ref{sec::TPC-H}.

\subsubsection{Aggregate functions}
Our firsts tests with aggregate functions, e.g. \texttt{avg()}, \texttt{sum()}
and \texttt{count()}, show in particular the impact of the filesystem cache of
the underlaying operating system.

All of the mentioned aggregate functions cause \textit{PostgreSQL} to perform a
sequential scan. The first query that was executed (\texttt{count(*)}) had a 
notably longer execution time compared to the next ones (\texttt{avg()} and
\texttt{sum()}). The query plans for all three queries are the same. After a
closer investigation of the logged activities of the operating system, we
identified the filesystem cache as the performance booster by caching the
database files that \textit{PostgreSQL} uses. In this case the kind of executing
the queries (single session vs. multi session) does not matter. The most
important setting is the fraction of main memory assigned to
\textit{PostgreSQL}: the lower the fraction the more main memory is available
for system caches.

\begin{figure}[H]
\centering
\includegraphics[width=\columnwidth]{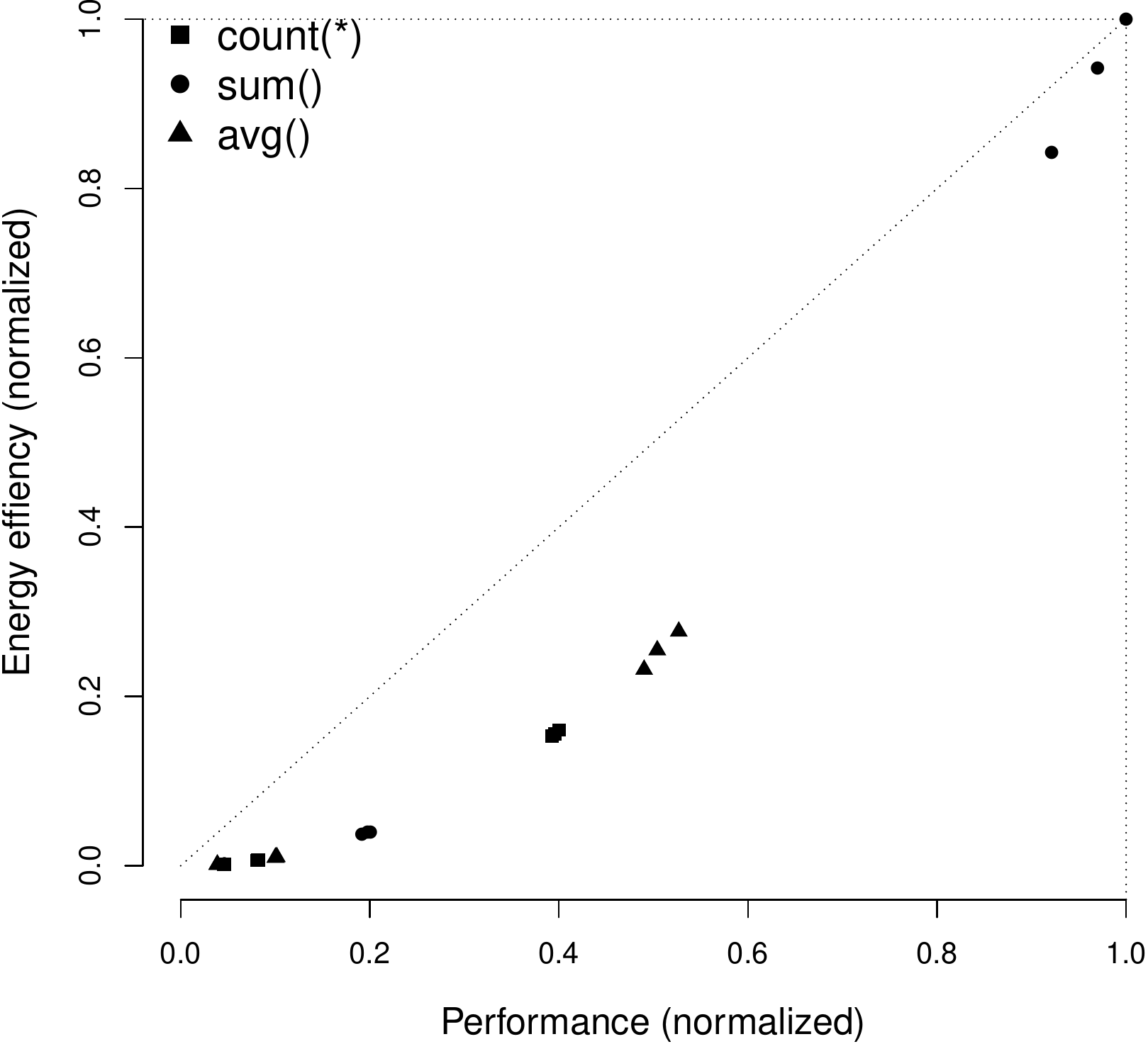}
\caption{Energy effiency for W22 group 1}\label{fig::w22-g1-normalized}
\end{figure}

The effect of file system caches provided by the ope\-ra\-ting system is
depicted in figure \ref{fig::w22-g1-normalized} for the mentioned aggregate
functions. Due to the longer execution time the energy consumption was higher
resulting in a lower energy effiency. Subsequent queries benefited of the
cache.

The suggested optimizations of the queries, e.g. the use of an indexed column
for \texttt{count()} (\texttt{count(column)} instead of \texttt{count(*)}), did
not change the execution times.

\subsubsection{Grouping and sorting}
The queries of group 2 (grouping) and group 3 (sorting) expands the
\texttt{count()} query from group 1 with SQL operators like \texttt{ORDER BY}
and \texttt{GROUP BY} to get comparable results. Those queries were composed to
study the impact of the mentioned SQL operators in reaction of the previous
tested benchmarks TCP-H and SSB.

The tests result indicates only a slight difference in terms of execution times
and energy consumption compared to the test results of group 1. The impact on
the energy consumption of the CPU and the main memory was not measurable. As a
result the calculated energy effiency remains the same.

\subsubsection{Selection of different data types}
According to our tests, the selection of a specific data type, e.g.
\texttt{VARCHAR}, \texttt{DATE} and \texttt{INTEGER}, has no impact on the query
execution time.

The other purpose of the queries of group 4 was to test the operators of the
different data types with and without having an index on the particular
table column. We observed a big impact on the execution time when an index was
used whereas the average energy consumption was slightly higher. The presence of
an index in combination of an operator\footnote{For example, the operators
greater than, less than and equals (\texttt{<}, \texttt{>} and \texttt{=}) are
valid for B-tree indices on numeric table columns in \textit{PostgreSQL}. There
are other index types, e.g. inverted index, for other column data types as well
as their specialized operators.} supporting this index lead to an immense
performance gain. Therefore the energy effiency is quite high compared to a
sequential scan. In contrast to the queries with an absent index, the queries
with an involved index showed an almost linear performance. This is crucial
since we used the largest table of TPC-H, \texttt{lineitem}, which also has the
greatest amount of rows throughout all queries of this group.

\textit{PostgreSQL} uses an index for a query when a) the index supports the
operator of the query and b) the costs for processing the index are lower than
sequentially scanning the table. Those costs are composed of customizable base
costs and dynamic cost estimations that \textit{PostgreSQL} gathers periodically
and statistically from all tables of a database.

We modified the settings for the base costs of loading and processing a database
and an index page (usually 8 KBytes of data) to favor the usage of indices,
e.g. if the database and indice files are stored on different storage devices
with different access speeds (the indice files are usually stored on the faster
one). Our test results remain the same because the data and index files are
stored on the same hard drive device.

\subsubsection{Joining tables and eleminating duplicates}
Based on our TPC-H and SSB benchmark results, we were interested in the
behaviour of \textit{PostgreSQL} dealing with table joins. There are two kinds
of joins supported in \textit{PostgreSQL}: unconditional and conditional joins.

Our first query of this W22 group deals with an unconditional join of two
tables where the cross product is further restricted by the conditions given
after the \texttt{WHERE} clause (one restriction per involved table). We
expected this query to be unperformant due to the cross product and the
successive restriction of the result set, but this was not the case. The query
plan reveals the (unintentional) use of an index for one restriction and a
sequence scan for the other one. So the results (performance, energy consumption
and the energy effiency) are the same as mentioned in the last section although
a sequence scan is part of the query plan.

\newpage
The other queries of this W22 group were composed to join two TPC-H tables
using inner\footnote{The SQL standard standardized an inner join as
\texttt{<table a> [INNER] JOIN <table b> WHERE <a.xyz> = <b.xyz>}.} and
equi-joins\footnote{An equi-join is in the form \texttt{<table a> JOIN <table b>
ON <a.xyz> = <b.xyz>}. The SQL standard allows shorthand for the column to
join by using the \texttt{USING} clause.} to examine differences in the query
plans.

Interestingly those queries indicate the same characteristics in terms of the
query planner. For \textit{PostgreSQL} it does not matter where the condition
for joining two tables is placed. In other words, \textit{PostgreSQL} does not
distinguish between an inner, implicit or equi-join.

Besides, the queries are composed to join two TCP-H tables with different
number of rows to investigate the join performance. The first two queries
joined the \texttt{lineitem} table with the \texttt{orders} and the
\texttt{part} table, respectivly. The last query joined the \texttt{orders}
with the \texttt{customers} table. Please refer to table \ref{tab::rownumbers}
to get an overview of the number of rows. Finally we select the amount of
joined rows by using the \texttt{count(*)} aggregate function. This forces
\textit{PostgreSQL} to use sequential scans for the mentioned tables.

\begin{table}[htb]
\centering\small
\begin{tabular}{|l|c|c|c|}
\cline{2-4} \multicolumn{1}{c|}{} & \multicolumn{3}{c|}{\textbf{Database size}}
\\
\hline \textbf{Table} & 1 GByte & 5 Gbyte & 10 GByte \\
\hline\hline \texttt{lineitem} & 6.001.215 & 299.999.795 & 599.986.052 \\
\hline \texttt{orders} & 1.500.000 & 7.500.000 & 15.000.000 \\
\hline \texttt{part} & 200.000 & 1.000.000 & 2.000.000 \\
\hline \texttt{customer} & 150.000 & 750.000 & 1.500.000 \\
\hline
\end{tabular}
\caption{Number of rows for some TPC-H tables}\label{tab::rownumbers}
\end{table}

As assumed, our test results indicate that the join performance is strongly
related to the row scan performance. The test results are similar to the ones
of our TPC-H and SSB benchmarks. This means they do not resemble much in their
energy consumption but in their execution times. Unsurprisingly the lower the
amount of rows to be scanned for joining, the lower is the execution time and
therefore the energy effiency is quite better. Although compared with query with
the unconditional join mentioned above, the performance is fundamentally worse.

We were also interested in the effects of eleminating duplicates from a result
set. For this purpose we formed a test query using the \texttt{DISTINCT()}
clause on a column of the \texttt{lineitem} table not having a supporting
index. The query plan revealed a sequence scan and the removal of duplicates by
hashing the values. Again, the test results showed the same characteristics as
all of our database tests performing a sequence scan.

% \subsubsection{W22 energy effiency}
% As 
% \begin{figure}[H]
% \centering
% \includegraphics[width=\columnwidth]{w22-effiency-connection-bw.pdf}
% \caption{Energy effiency for W22 workload}\label{fig::w22}
% \end{figure}

\section{Summary}
At first, our tests with our database server using normal HDDs indicates an
energy increase of roughly 9 W when the HDDs are fully utilized. Compared to the
energy consumption of the other measured core components during the tests, this
is insignificant. The argument, SSDs should be prefered because they consume up
to 12 times less energy compared to HDDs, is invalid in this context.

As \textit{Lang et all.} stated in the summary of \cite{/8/}, evaluating the
energy effiency of a DBMS needs the inclusion of entire workloads, not just
single queries. This study makes use of three different and complete workloads
that allows are more comprehensive look at the energy effiency of a relational
DBMS. Most of  the benchmark queries caused a massive usage of sequential scans.
This implies that the sequential read performance is an extremly important
factor that affects the energy consumption. Actual SSDs clearly outperform
normal HDDs but in this case enterprise grade HDDs can be used because they
offer nearly the same performance as SSDs.

As mentioned in the introductionary section of this paper, there are more
factors and not only technical parameters that influence the performance and
thus the energy effiency of a database server.

For example, the filesystem cache provided by the ope\-ra\-ting system is more
relevant for the execution of a database query in \textit{PostgreSQL} than its
internal cache. Based on our experiments, we recommend not to assign more than
50 percent of the main memory to \textit{PostgreSQL} for operations. With more
assigned main memory the remaining processes of the operating system are forced
to use the remaining portion. This causes the operating system to swap this
portion to the hard drive which leads to a dramatic reduction of the
performance.

Another important factor for the energy effiency of the used database benchmark
is the kind of accessing \textit{PostgreSQL} (single vs. multiple database
connections). Our assumption, subsequent queries of the benchmarks would benefit
from \textit{PostgreSQL}'s internal cache by using just a single database
connection, does not come true. In fact, the opposite performed better.

Our tests also indicate the fact that energy saving settings are
counterproductive for a database server that is reasonably utilized because it
decreases the overall system performance.

\section{Limitations and future work}
Since our tests were based on \textit{PostgreSQL} as the DBMS, the results can
not be purely adopted to other enterprise data management systems.
\textit{PostgreSQL} was chosen because its source code is freely available and a
good starting point for academic research. However, the basic database
technologies like indices and data access patterns can be found in almost any
other DBMS. Currently a newer version of
\textit{PostgreSQL}\footnote{\textit{PostgreSQL} version 9.2.3} is available.
Its release notes announces a better usage of indices and in general a higher
performance. In the future all tests could be redone to analyze the impact of
software improvements on the energy effiency.

Another limitation on this was the strict use of HDDs. While a number of studies
have reported the optimization in performance by the use of SSDs, this study
considered only the use of HDDs as a basis since these are still common 
in datacenter operations. A future research task could consider a benchmarking
scenario making a hybrid use of SSDs and HDDs where SSDs could be used to
improve access to indices and HDDs for the storage of the raw data.

\bibliography{paper2}
\bibliographystyle{abbrv}

\appendix
\section{Measurement arrangement\newline and
instrumentation}\label{sec::testarrangement}
\begin{figure}[htb]
\centering
\subfigure[\label{subfig::ATX}ATX]{\includegraphics[width=.8\columnwidth]
{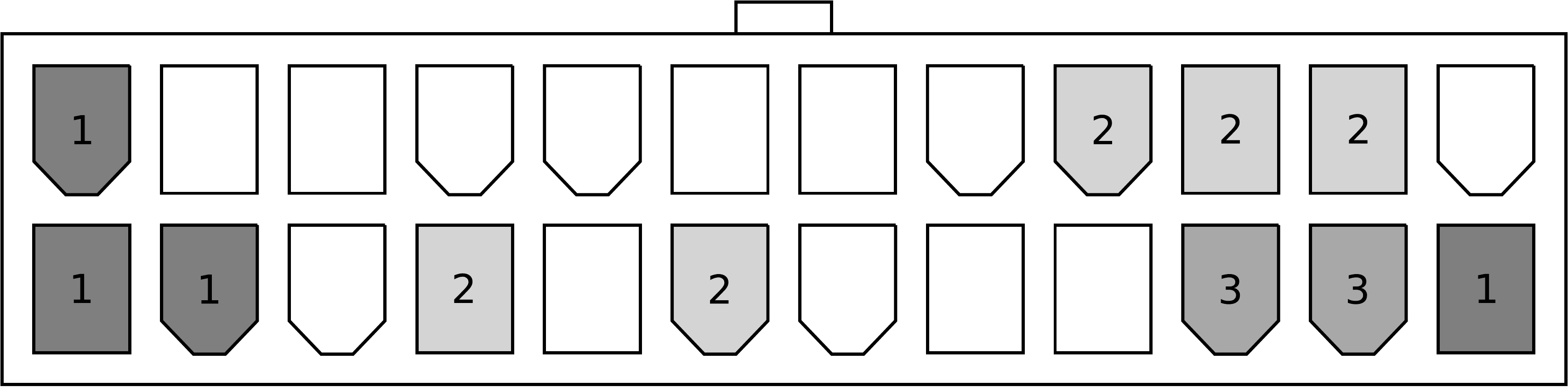}}
\subfigure[\label{subfig::ATX12}ATX12]{\includegraphics[width=.2\columnwidth]
{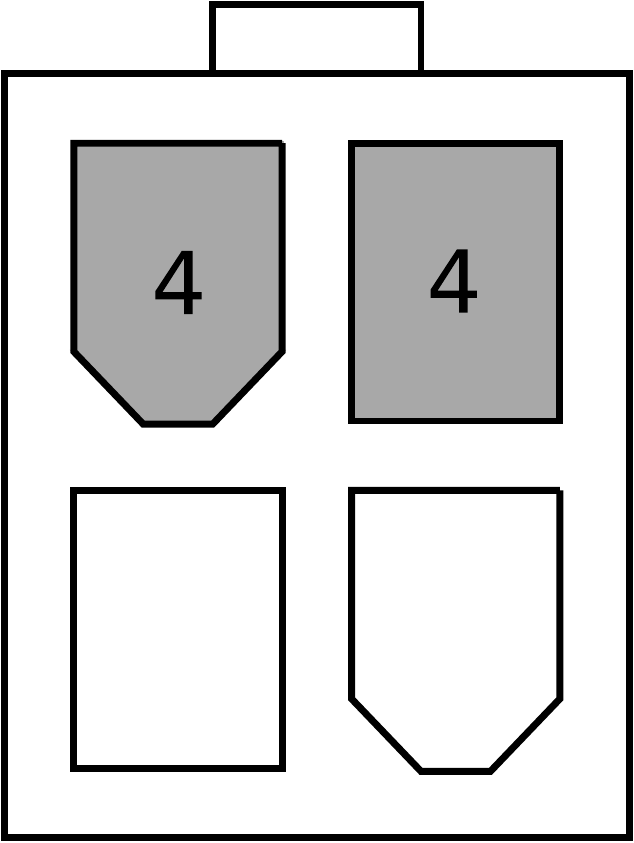}}
\hspace{.2\columnwidth}
\subfigure[\label{subfig::molex}Molex]{\includegraphics[width=.25\columnwidth]
{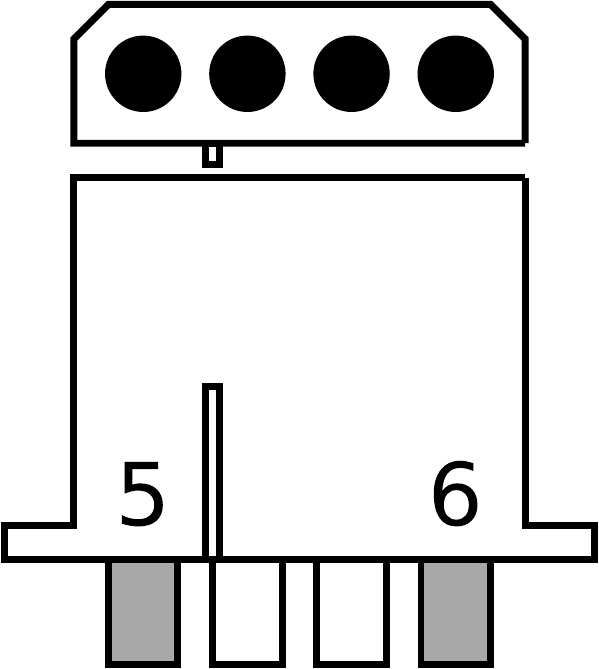}}
\caption{Pins used for measurements}\label{fig::lanes}
\end{figure}

To measure the energy consumption of the mainboard we modified the ATX power
cord as depicted in figure \ref{subfig::ATX}: the wires of related pins were
multiplexed into single cables and demultiplexed on the opposite side to its
original configuration. The purpose is to have an easier and accurate way 
measuring the voltage drop of these cables.

The same applies for the energy consumption of the CPU where the additional
ATX12 cable was modified as shown in figure \ref{subfig::ATX12}. To measure the
energy consumption of the two hard drives in the RAID array we used a
\textit{Molex} Y-splitter adapter to get the combined consumption and modified
it respectivly. This is shown in figure \ref{subfig::molex}.

In general, figure \ref{fig::lanes} shows the pin usage. Related pins are
grouped into pin groups which are represented by a number between 1 and 6. Each
group stands for a power lane to be measured. Table \ref{tab::lanes} shows the
power lanes and the pin group.

To measure the power consumption of the mentioned components, six digital multi
meters (\emph{Uni-T UT61-B} with an average accuracy of $\pm$ 0.8 percent) and
an energy logger (\emph{Voltcraft energy logger 4000} with an accuracy $\pm$ 1
percent) for the overall power consumption were applied. The benefit of this
test arrangement is a very accurate measurement compared to other techniques
such as power clamp meters.

\begin{table}[htb]
\centering\small
\begin{tabular}{|l|l|p{.3\columnwidth}|c|}
\hline \textbf{Name} & \textbf{Voltage} & \textbf{provides} &
\textbf{Pin group}\\
\hline \multicolumn{4}{c}{\emph{Main power supply (ATX)}} \\
\hline 3.3V & +3.3 V & Mainboard (main memory) & 1\\
\hline 5V & +5 V & Mainboard & 2 \\
\hline 12V1 & +12 V & Mainboard & 3 \\
\hline \multicolumn{4}{c}{\emph{ATX12 connector}} \\
\hline 12V2 & +12 V & CPU & 4 \\
\hline \multicolumn{4}{c}{\emph{Molex adapter}} \\
\hline & +5 V & internal drives & 5 \\
\hline & +12 V & internal drives & 6 \\
\hline
\end{tabular}
\caption{Power lanes}\label{tab::lanes}
\end{table}
%\balancecolumns
\end{document}